\documentclass[aps,english,pre,preprint,superscriptaddress,showpacs]{revtex4-1}

\usepackage{graphicx}
\usepackage{amssymb}
\usepackage{babel}
\usepackage{color}
\usepackage{multirow}
\usepackage[T1]{fontenc}
\usepackage{subfigure}

\begin{document}
\title{\textbf{Testing the generalized conjugate field formalism in the kinetic Ising model with nonantisymmetric magnetic fields: A Monte Carlo simulation study }}

\author{Y. Y\"{u}ksel}
\affiliation{Dokuz Eylul University, Faculty of Science, Physics Department, 
Tinaztepe Campus, 35390 Izmir, Turkey}

\date{\today}
%\pacs{pac1,pac2}

\begin{abstract}
We have performed Monte Carlo simulations for the investigation of dynamic phase transitions on a honeycomb lattice which has garnered a significant amount of interest from the viewpoint of tailoring the intrinsic magnetism in two-dimensional materials. For the system under the influence of time-dependent magnetic field sequences exhibiting the half-wave anti-symmetry, we have located a second order dynamic phase transition between dynamic ferromagnetic and dynamic paramagnetic states. Particular emphasis was devoted for the examination of the generalized conjugate field  formalism 
previously introduced in the kinetic Ising model [\color{blue}Quintana and Berger, Phys. Rev. E \textbf{104}, 044125 (202); Phys. Rev. E \textbf{109}, 054112] \color{black}. Based on the simulation data, in the presence of a second magnetic field component with amplitude $H_{2}$ and period $P/2$, the half-wave anti-symmetry is broken and the generalized conjugate field formalism is found to be valid for the present system. However, dynamic scaling exponent significantly deviates from its equilibrium value along with the manifestation of a dynamically field polarized state for non-vanishing $H_{2}$ values.
\end{abstract} 

\maketitle

\section{Introduction}\label{intro}
Response of a ferromagnet to the presence of time-dependent and oscillating magnetic field in the context of dynamic phase transition (DPT) (a phenomenon of statistical physics emerging in ferromagnets below their ordering temperature $T_{c}$) attracted a significant amount of attention both in theoretical and experimental grounds \cite{tome,lo,yuksel1,berger,robb_2}.
 In the presence of a single-component sinusoidal field of period $P$ and amplitude $H_{0}$, time series of instantaneous magnetization $m(t)$ can either follow the external perturbation with a small amount of delay or $m(t)$ will be unable to follow the periodic field sequence and it tends to oscillate around some non-zero magnetization. This scenario is governed by a competition mechanism between two time scales, namely, the relaxation time $\tau$ of the system and field period \cite{chakrabarti}. In the fast critical dynamics regime $(P\ll\tau)$, the system attains dynamic ferromagnetic phase where $m(t)$ cannot follow the fast reversal of magnetic field whereas in the slow critical dynamics regime $(P\gg\tau)$, the system stays in the dynamic paramagnetic regime with a small phase lag between $m(t)$ and $H(t)$ sequences. This dynamic phase transition is associated to a dynamic symmetry breaking phenomenon \cite{riego_2}.

In the absence of any constant bias field $H_{b}$, DPT universality class is found to be identical to that observed in equilibrium thermodynamic phase transitions (TPT) with the same critical exponents \cite{korniss,buendia}. Besides, in the presence of non-zero $H_{b}$ term, it was found that the bias field plays the role of the conjugate field $H^{*}$ of the dynamic order parameter when the field sequence is in conventional sinusoidal or square wave form. This is analogous to the relation between magnetization $M$ and longitudinal magnetic field $H$ in the equilibrium thermodynamic system. Nevertheless, the similarities between DPT and TPT must be handled with caution, as some prominent differences may appear upon introducing a constant bias term $H_{b}$. The so-called 
metamagnetic anomaly phenomenon is especially observed in the slow critical
dynamics regime, and it appears  as multiple-symmetric peaks in the $H_{b}$ dependence of scaled variance $\chi_{Q}$ associated to the dynamic order parameter $Q$. Note that although the metamagnetic anomalies have been observed in both theoretical \cite{buendia2,mendes} and experimental studies \cite{riego,ramirez}, there is not any equilvalence of this behavior for a regular equilibrium ferromagnet where the magnetic susceptibility versus magnetic field curve exhibits a broad maxima centered around zero field \cite{berger_eq,mcKenzie}.

Generally, the vast majority of the recent and preceding literature dealing with DPT phenomenon have  considered time-dependent magnetic field $H(t)$ sequences exhibiting the property of half-wave anti-symmetry defined by $H(t)=-H(t+P/2)$. This anti-symmetry in $H(t)$ is valid for sinusoidal and square wave forms of field oscillations. In the dynamic paramagnetic regime, half-wave anti-symmetry is also observed in the time series of instantaneous magnetization obeying $m(t)=-m(t+P/2)$, consequently resulting in $Q=0$. However, it is possible to break the half-wave anti-symmetry by introducing a second harmonic contribution with amplitude $H_{2}$ and period $P/2$. Average dynamic order parameter will now be non-zero even in the dynamic paramagnetic regime. In such systems, bias field no more corresponds to the conjugate field of $Q$. In this regard, concept of generalized conjugate field $H^{*}$ for any general field sequence $H(t)$ has been recently introduced by Quintana and Berger \cite{quintana,quintana2}, guaranteeing the condition $Q(H^{*})=-Q(-H^{*})$ for arbitrary $P$ and $H_{0}$ parameters. As a limiting case, for the single-component sinusoidal magnetic fields, it has been found that $H_{b}=H^{*}$ holds true, reinforcing the prior statement that 
the bias field plays the role of the conjugate field $H^{*}$ of the dynamic order parameter.

It has been theoretically and experimentally verified that the generalized conjugate field $H^{*}$ can be given by \cite{quintana,quintana2}
\begin{equation}\label{eq1}
H^{*}=H_{b}+\Delta H(H_{b},H_{2}),	
\end{equation}
where the deviation $\Delta H$ is defined as 
\begin{equation}\label{eq2}
	\Delta H=-\frac{1}{2}\left[H_{b}(Q)+H_{b}(-Q)\right],
\end{equation}
corresponding to a generalized field sequence 
\begin{equation}\label{eq3}
	H(t)=H_{b}+H_{0}\sin\left(\frac{2\pi t}{P}\right)+H_{2}\sin\left(\frac{4\pi t}{P}\right), 
\end{equation}
which is composed of a constant bias term $H_{b}$, a fundamental part with amplitude $H_{0}$ and period $P$, and a third component of period $P/2$ and amplitude $H_{2}$. It is worth to mention that very recently, kinetic Ising model under a non-antisymmetric magnetic field has been investigated using Monte Carlo simulations \cite{vatans}. Ising universality has been found to be conserved in the absence of half-wave antisymmetry for small $H_{2}$ values. 
However, it has also been reported that further analysis are required for the complete understanding of universality properties, as well as for the validity of the conjugate-field formalism for larger values of the amplitude of the second harmonic field component $H_{2}$.

On the other hand, honeycomb lattice structure is often utilized in calculations regarding the material science and in the two-dimensional magnetism research. After the exfoliation of graphene in 2004 \cite{novoselov}, exploration of magnetism and  DPT properties in two-dimensional (2D) materials has gained particular interest \cite{huang,huang2,vatansever_e,yuksel_0}. Among these works, van der Waals crystals of monolayer chromium trihalides \cite{webster,xue} and ruthenium halides \cite{huang2,sari} are remarkable ferromagnetic materials exhibiting intrinsic magnetism with Curie points $T_{c}=45K$ $(\mathrm{CrI_{3}})$ \cite{huang} and $T_{c}=14.2K$ $(\mathrm{RuCl_{3}})$ \cite{sari}. According to the crystal structure analysis of these materials, magnetic atoms are arranged in a honeycomb lattice formation. 

From the perspectives of DPT and 2D magnetism, our aim is to examine the recent developments such as applicability of generalized conjugate field formalism and related universality aspects with larger values of $H_{2}$ component to kinetic Ising model defined on a honeycomb lattice.
For this aim, we have calculated the critical period at which a DPT takes place between dynamic ferromagnetic and dynamic paramagnetic states. We have also examined the scaling behavior of $Q$ as a function of $H^{*}$ by estimating the critical exponent $\delta$ for non-zero $H_{2}$ values. 
The organization of the article is as follows: In Sec. \ref{formulation}, we present our model and simulation details. Sec. \ref{results} is devoted for numerical results and related discussions. Finally Sec. \ref{conclude} contains our conclusions.

\section{Model and Formulation}\label{formulation}
The system is simulated on a honeycomb lattice with coordination number $z=3$ based on the following atomistic spin Hamiltonian,  
\begin{equation}\label{eq4}
	\mathcal{H}=-J\sum_{\langle ij \rangle}S_{i}S_{j}-H(t)\sum_{i}S_{i},
\end{equation}
where $J>0$ is the nearest-neighbor ferromagnetic exchange coupling and $S_{i}=\pm 1$ is the pseudo-spin variable located on the site $i$. The first summation in Eq.(\ref{eq4}) is calculated over the nearest-neighbor spin couples whereas the last term is over all the lattice sites.  $H(t)$ represents a time-dependent magnetic field which is defined by Eq. (\ref{eq3}). Note that the complete field signal is spatially uniform.
Numerical calculations have been carried out within the framework of Monte Carlo simulations with Metropolis local spin updating scheme \cite{newman,binder}. We empose periodic boundary conditions in each direction. The lattice points have been randomly swept on a lattice with linear dimension $L$ ranging from $64$ to $256$. In order to calculate the physical quantities, $1.1\times10^{4}$ period cycles of the oscillating field were considered and the initial $10^{3}$ cycles were discarded for thermalization. Hence, the computational time need for measuring the quantities depends on the period $P$. In the absence of any magnetic bias field term, we have considered up to 500 independent realizations at a constant temperature. The error bars have been evaluated using Jackknife method \cite{newman} for which the data set containing 500 individual measurements for each quantity have been divided into 20 subgroups. In this regard, the obtained error bars were observed to be smaller than the data points. In order to guarantee that the system stays in the multi-droplet regime \cite{sides_1, rikvold_1}, we fixed the temperature at $T=0.8T_{c}$ where $T_{c}/J=1.519$ is the thermodynamic phase transition temperature of honeycomb lattice \cite{fisher,fisher_book}.

Using the time series of magnetization, dynamic order parameter in the $k^{th}$ cycle can  be calculated via
\begin{equation}\label{eq5}
Q(k)=\frac{1}{(2t_{1/2})}\int_{(k-1)(2t_{1/2})}^{k(2t_{1/2})}m(t)dt,	
\end{equation}	
where $t_{1/2}$ denotes the half-period (i.e. $P=2t_{1/2}$) of the oscillating magnetic field.
From Eq. (\ref{eq5}), average dynamic order parameter $\langle Q\rangle$ can also be evaluated by performing the averaging procedure over successive cycles of $H(t)$. Besides, in order to locate the critical period, we also calculate the dynamic scaling variance of $Q$ using
\begin{equation}\label{eq6}
\chi_{Q}=N\left[\left\langle Q^{2}\right\rangle_{L}-\langle Q\rangle^{2}_{L}\right].	
\end{equation}	
Precise location of the critical period at which a phase transition occurs between the dynamic ferromagnetic and paramagnetic states can be determined by Binder cumulant analysis \cite{binder0} which allows us to benefit from the higher order moments of $Q$ according to
\begin{equation}\label{eq7}
V_{L}=1-\frac{\langle Q^{4}\rangle}{3 \langle Q^{2}\rangle}.
\end{equation}	
The intersection point of $V_{L}$ curves correspond to the critical point of the system. Note that we set $k_{B}=1$, and scale the field amplitudes in units of $J$ for the sake of convention in our calculations. 

\section{Results and Discussion}\label{results}
\begin{figure}[!h]
	\center
\includegraphics[width=7.0cm]{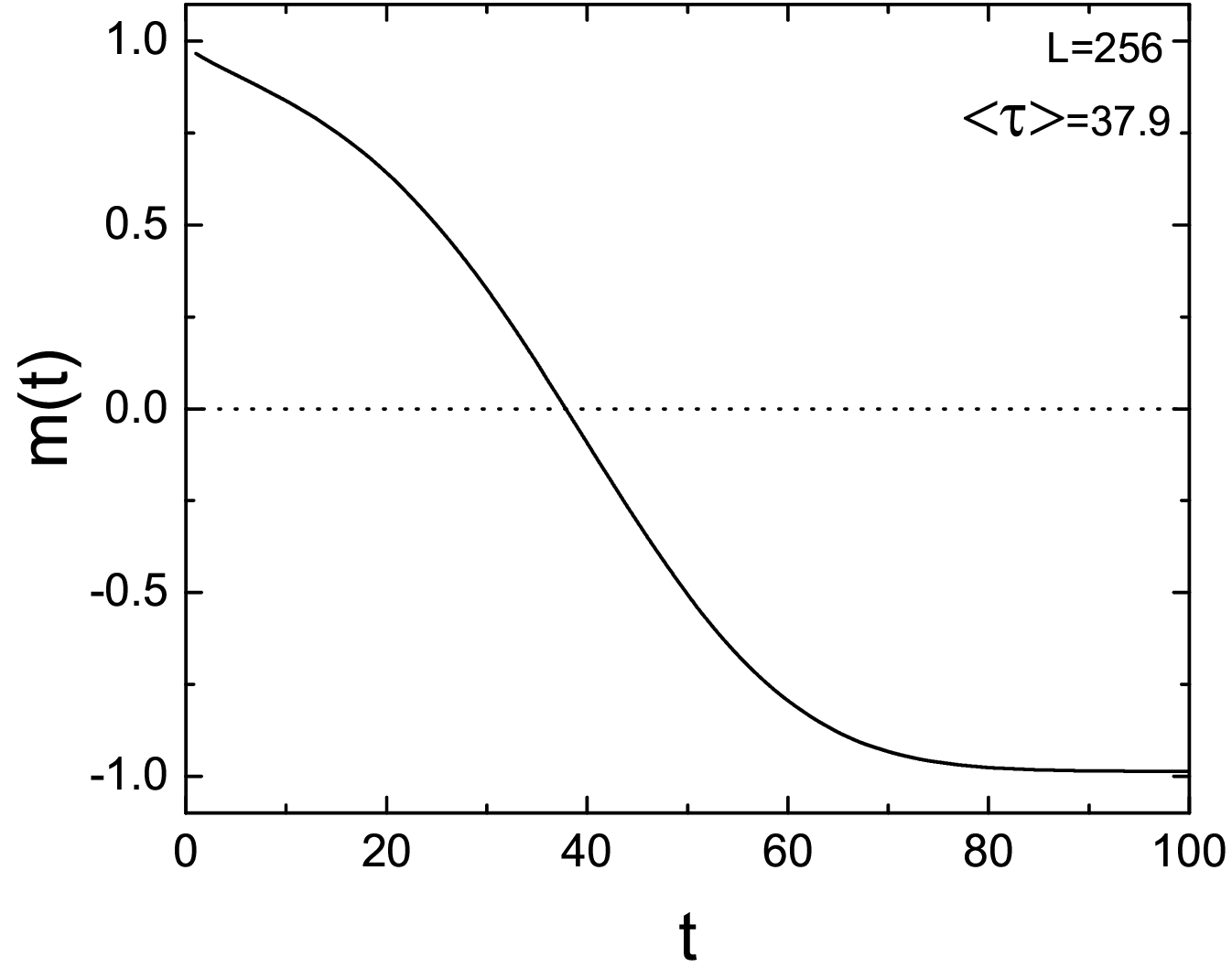}
	\caption{Variation of $m(t)$ as a function of time $t$. Numerical data was collected for a lattice with $L=256$ and at $T=0.8T_{c}$ with a constant bias field $H_{b}/J=-0.3$. The crossing point of the horizontal dotted-line and calculated curve gives an estimate for the relaxation time $\tau$ of the system.}\label{fig1}
\end{figure}

In order to quantify the competition mechanism between period $P$ and relaxation time $\tau$ which has been briefly summarized in Sec. \ref{intro}, we introduce the following relation \cite{park}
\begin{equation}\label{eq8}
	\Theta=\frac{t_{1/2}}{\left\langle\tau\right\rangle},
\end{equation}
where $\left\langle\tau\right\rangle$ denotes the average relaxation time  which is defined as the time needed for $m(t)$ to vanish momentarily in a dynamic system for which initially all spins are aligned anti-parallel to the external magnetic field. We have estimated $\left\langle\tau\right\rangle$ for the present honeycomb lattice by setting all spins pointing in the anti-parallel direction with respect to a constant bias field $H_{b}/J=-0.3$ and we inspect the time series of $m(t)$ sequence. According to our numerical data, we find $\left\langle\tau\right\rangle=37.9$ (e.g., see Fig. \ref{fig1}) which can be compared to the results obtained for decorated triangular ($\left\langle\tau\right\rangle=29.6$) \cite{yuksel}, kagome \cite{vatansever_zd} ($\left\langle\tau\right\rangle=55.8$), and square ($\left\langle\tau\right\rangle=74.6$) \cite{robb_1,sides_2} lattices. Note that relatively large value obtained for square lattice follows from the fact that Glauber-type single-spin-flip
algorithm generally produces over-estimated numerical values in comparison to those obtained using Metropolis dynamics. 

\begin{figure*}[!h]
	\center
	\subfigure[\hspace{0cm}] {\includegraphics[width=7.0cm]{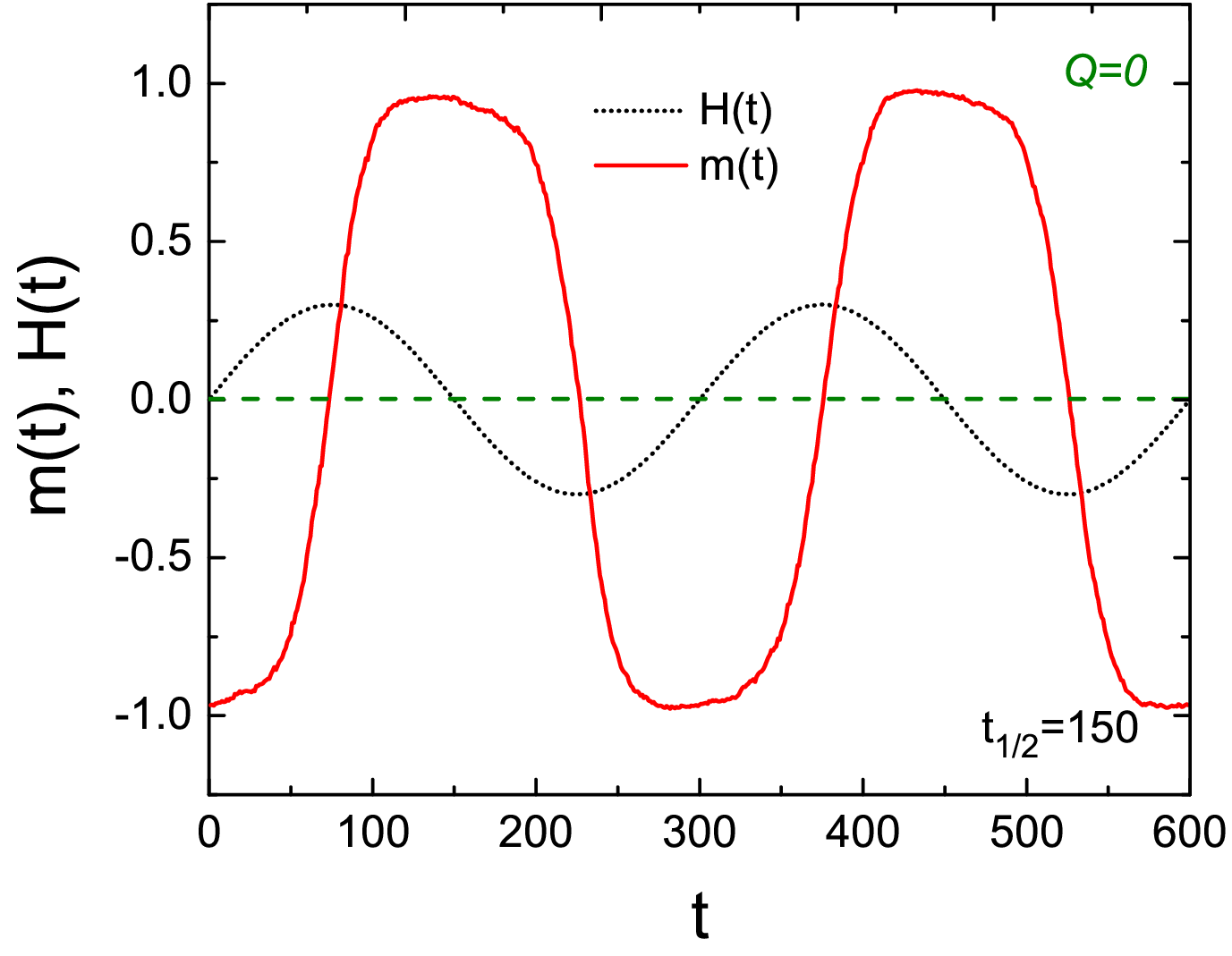}}
	\subfigure[\hspace{0cm}] {\includegraphics[width=6.8cm]{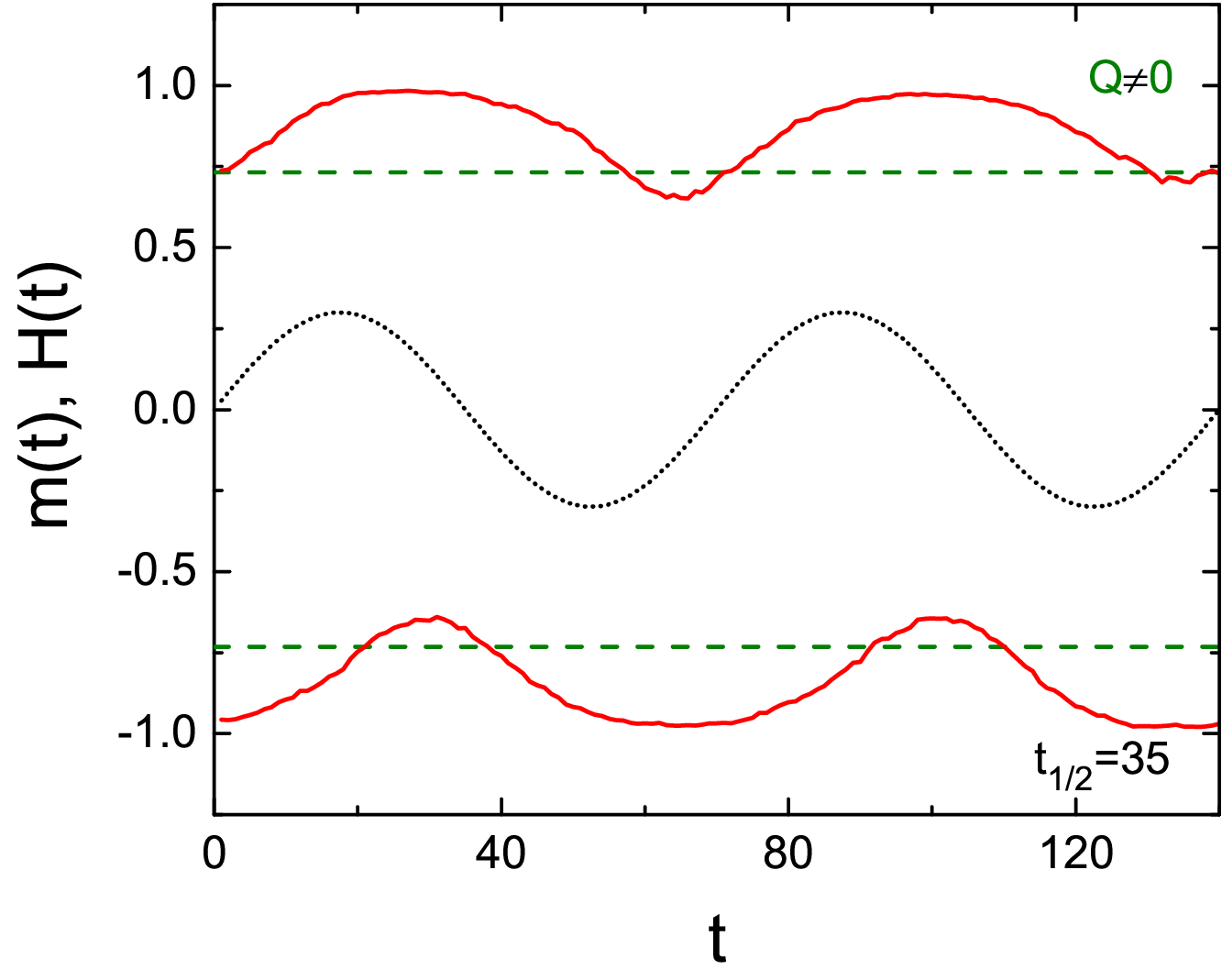}}\\
	\subfigure[\hspace{0cm}] {\includegraphics[width=7.0cm]{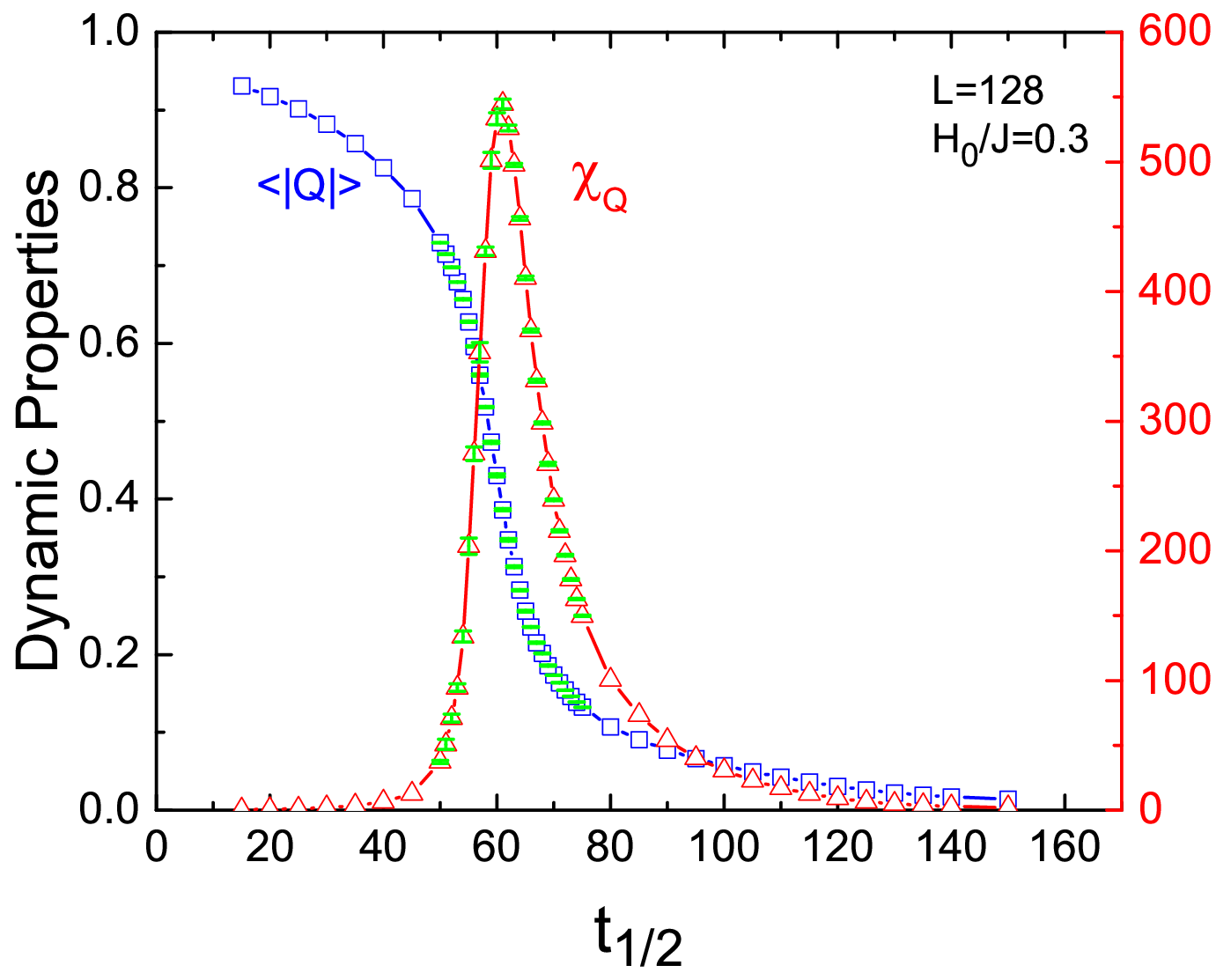}}
	\subfigure[\hspace{0cm}] {\includegraphics[width=7.0cm]{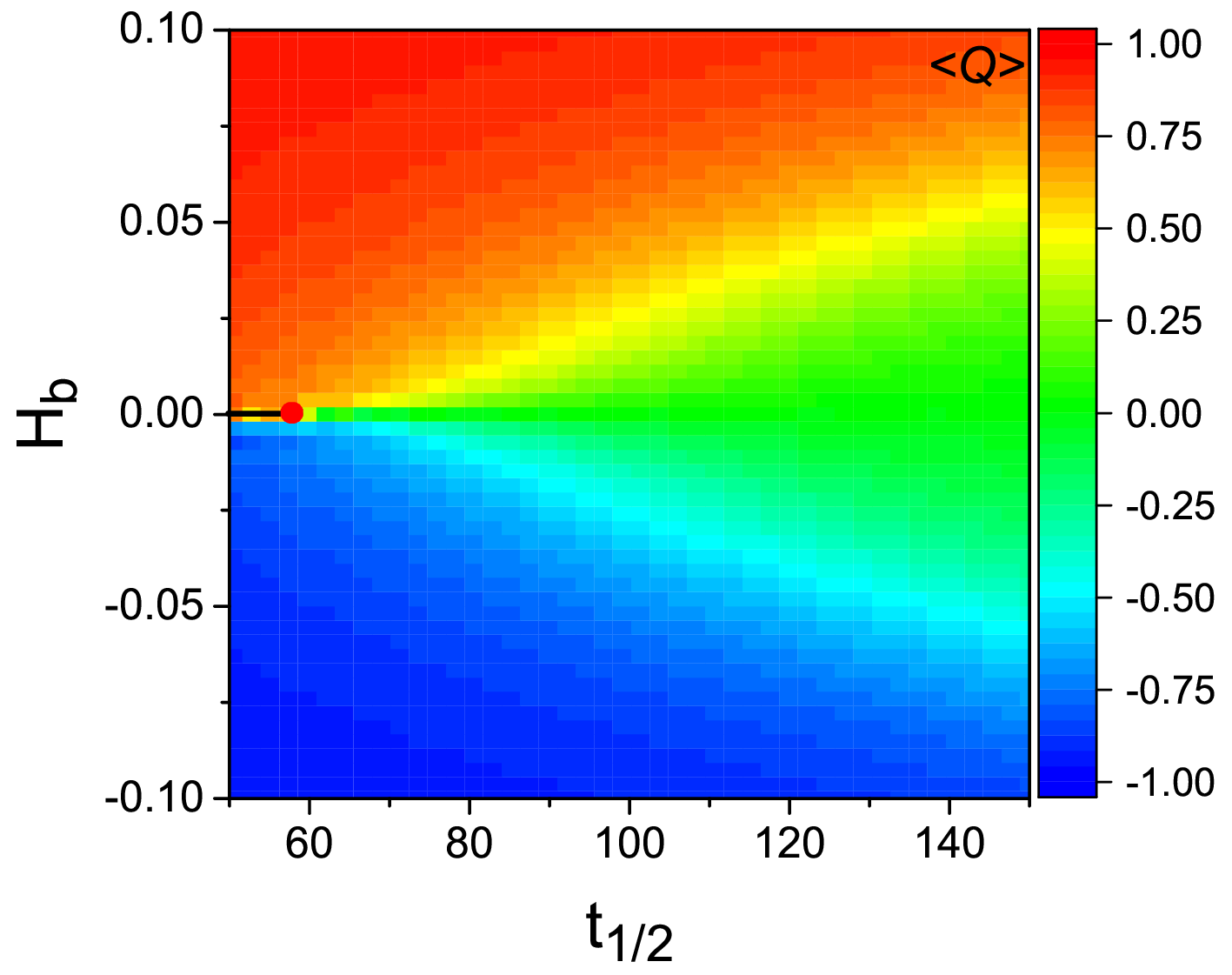}}
	\caption{Time series of $m(t)$ (solid red lines) and $H(t)$ (solid black lines) curves in (a) dynamic paramagnetic and (b) dynamic ferromagnetic regimes. (c) Variations of dynamic order parameter $\langle |Q| \rangle$ and dynamic scaling variance $\chi_{Q}$ with half-period $t_{1/2}$. (d) 2D contour plot of $Q$ as a function of $t_{1/2}$ and bias field $H_{b}$. The location of critical point is denoted by filled red circle. All calculated properties have been evaluated at $T=0.8T_{c}$  with $H_{0}/J=0.3$ and  $H_{2}=0.0$.}\label{fig2}
\end{figure*}

Figs. \ref{fig2}a and \ref{fig2}b illustrate representative $m(t)$ curves calculated for  $(H_{b},H_{2})=0.0$, corresponding to dynamic paramagnetic  $(P> P_{c})$ (c.f. Fig. \ref{fig2}a) and dynamic ferromagnetic $(P<P_{c})$ (c.f. Fig. \ref{fig2}b) states. From Fig. \ref{fig2}a, we see that $m(t)$ exhibits periodic oscillations with respect to the sinusoidally alternating magnetic field, indicating that the slow critical dynamics regime is manifested whereas from Fig. \ref{fig2}b, it can be inferred that $m(t)$ does not exhibit full reversal but oscillates around some non-zero value reinforcing the fast critical dynamics regime. Note that the dynamic ferromagnetic regime is two-fold  degenerate with non-zero $\pm Q_{0}$ (see the dashed lines in Fig. \ref{fig2}b). It is also important to notice that the applied field sequence shown in Fig. \ref{fig2} containing only a fundamental sinusoidal
component with period $P$ and amplitude $H_0$ exhibits the half-wave anti-symmetry property resulting in $Q\approx0$ in the dynamic paramagnetic regime.  In case of field sequences featuring the half-wave anti-symmetry (i.e., for $(H_{2},H_{b})=0)$, average dynamic order parameter $\langle |Q|\rangle$ and the corresponding scaled variance $\chi_{Q}$ obey the following scaling relations  at the dynamic critical period $(P_{c}=2t_{1/2}^{c})$ 
\begin{equation}\label{eq9}
\langle |Q|\rangle\propto L^{-\beta/\nu}	
\end{equation}
\begin{equation}\label{eq9}
	\chi_{Q}\propto L^{\gamma/\nu}.	
\end{equation}
The critical exponent ratios $\beta/\nu$ and $\gamma/\nu$ are found to be the same for DPT and TPT cases within some statistical error \cite{buendia}. From Fig. \ref{fig2}c, we find that $\langle |Q|\rangle$ exhibits a second order dynamic phase transition and the corresponding dynamic scaling variance exhibits a divergent behavior around $t_{1/2}^{c}$. Moreover, for a field sequence $H(t)$ with $H_{2}=0.0$, bias field $H_{b}$ corresponds to the conjugate field of $Q$. This can bee seen from Fig. \ref{fig2}d where we present the contour plot of $\langle Q\rangle$ in $(t_{1/2},H_{b})$ plane. It can be deduced from this figure that in the dynamic ferromagnetic phase $(t_{1/2}<t_{1/2}^{c})$, a first order phase transition (i.e. sharp transitions between blue and red regions) emerges between two degenerate ground states when crossing $H_{b}=0.0$ line. The approximate location of critical half period is marked by the filled red circle. For larger periods, a dynamic paramagnetic phase emerges with evident side bands (or meta-magnetic anomalies) characterized by the green triangular region in the ($t_{1/2} \ \mathrm{vs} \ H_{b}$) phase space.

\begin{figure}[!h]
	\center
	\includegraphics[width=8.0cm]{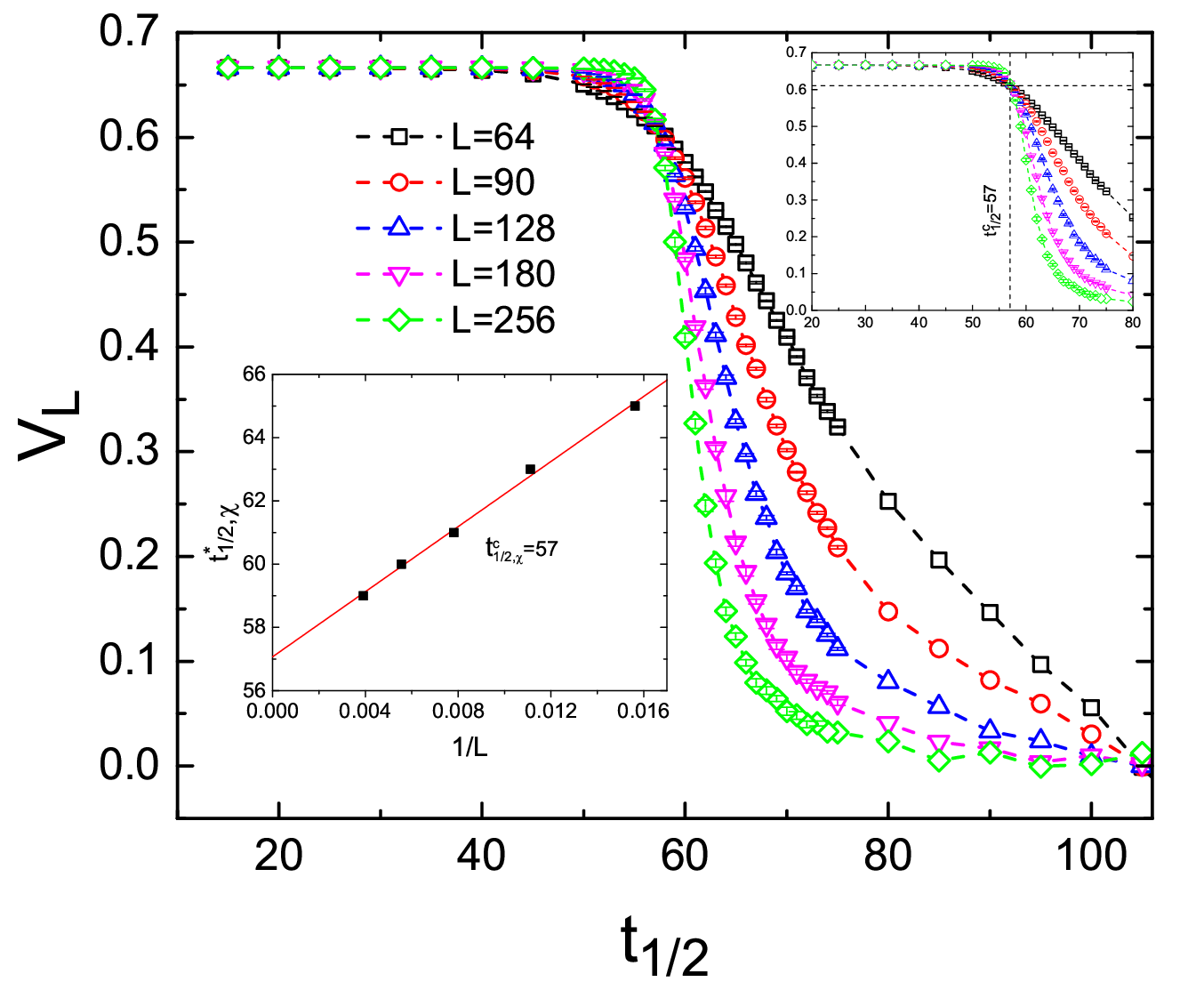}
	\caption{Binder cumulant curves $V_L$ as functions of half period $t_{1/2}$. The upper-right inset shows the critical region wheres the lower-left inset illustrates the critical half-period $t_{1/2}^{c}$ obtained from the finite-size scaling analysis.}\label{fig3}
\end{figure}

In order to precisely locate the dynamic phase transition point $t_{1/2}^{c}$, we have calculated the Binder cumulant $V_{L}$ given by Eq. (\ref{eq7}) for different lattice sizes. The crossing point of the curves with different system sizes shows the precise location of the critical half-period $t_{1/2}^{c}$ (see Fig. \ref{fig3}). Our simulation results suggest that $t_{1/2}^{c}=57$ for the present system. The horizontal line in the upper inset marks the universal value of the cumulant $V_{L}^{*}=0.6106924(16)$ of the Ising model in 2D obtained at the critical point \cite{kamieniarz,selke,salas}. In order to verify and reinforce the obtained value of $t_{1/2}^{c}$, we have also performed finite-size scaling analysis for the scaled variance $\chi_{Q}$ which is defined by Eq. (\ref{eq6}). The results are plotted as an inset in Fig. \ref{fig3} in which the pseudo-critical half period $t_{1/2}^{*}$ values obtained from $\chi_{Q}$ peak as a function of $1/L$ shows a linear variation with the slope $t_{1/2}^{c}=57$.

\begin{figure*}[!h]
	\center
	\subfigure[\hspace{0cm}] {\includegraphics[width=7.5cm]{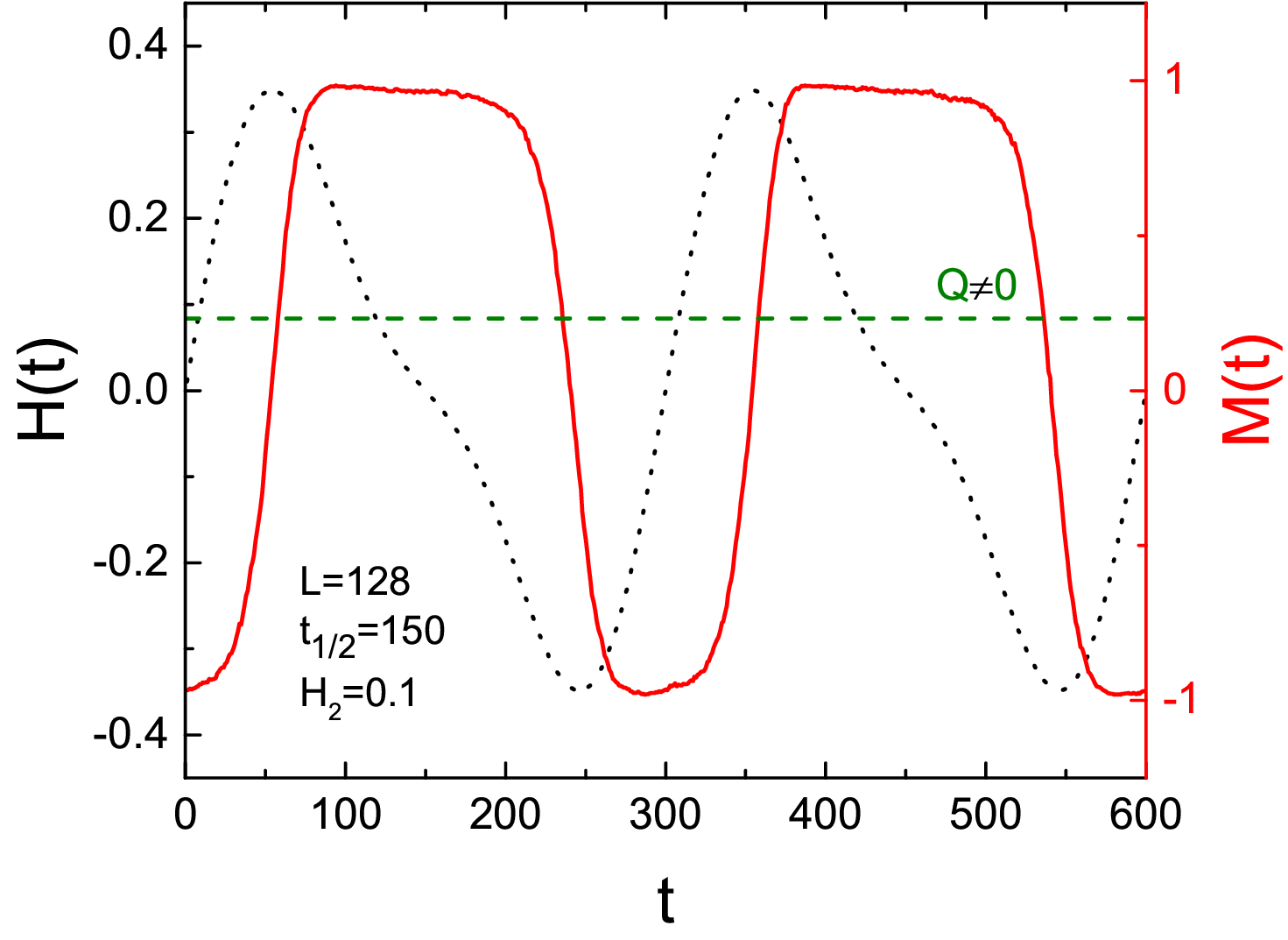}}
	\subfigure[\hspace{0cm}] {\includegraphics[width=7.0cm]{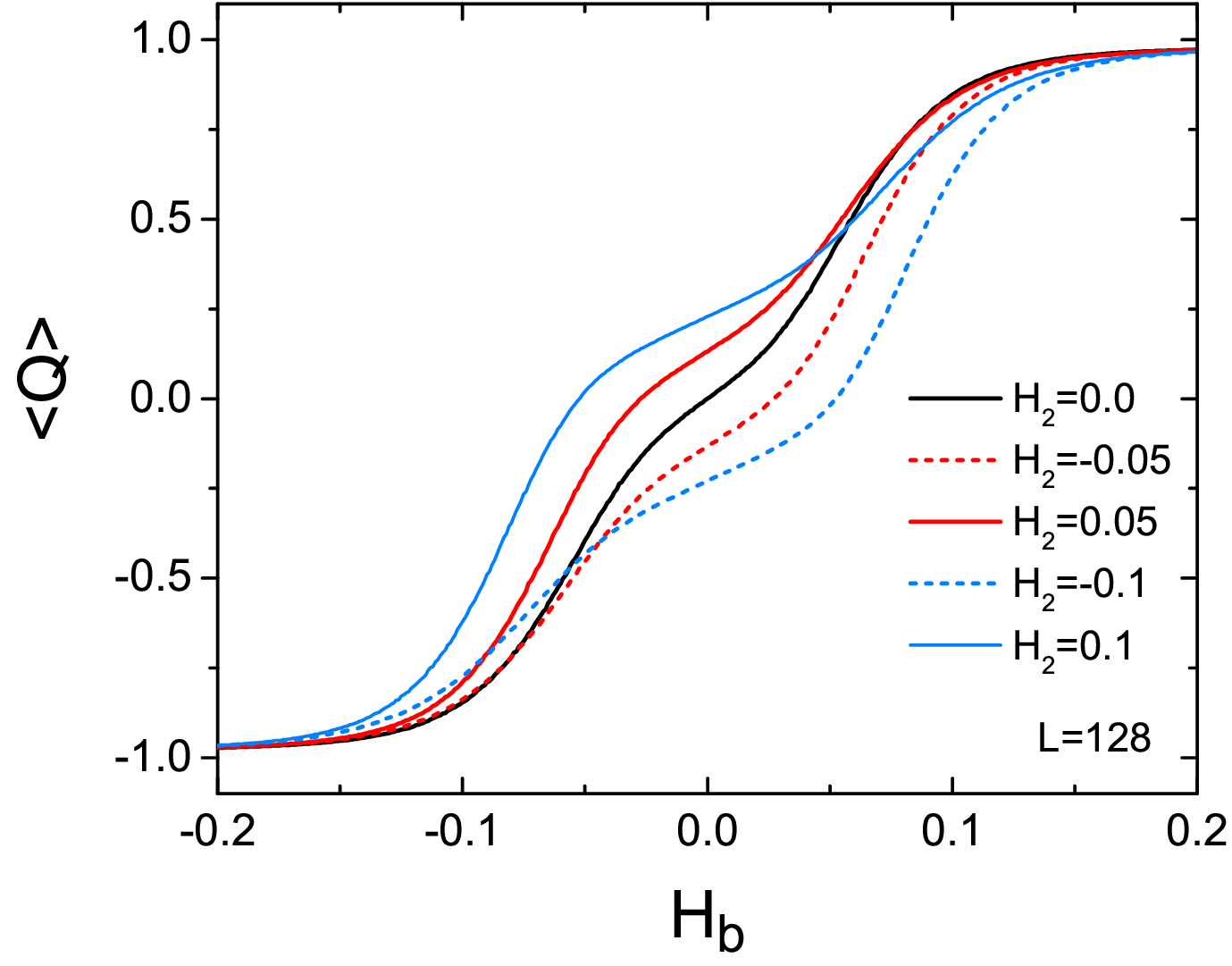}}\\
	\subfigure[\hspace{0cm}] {\includegraphics[width=7.2cm]{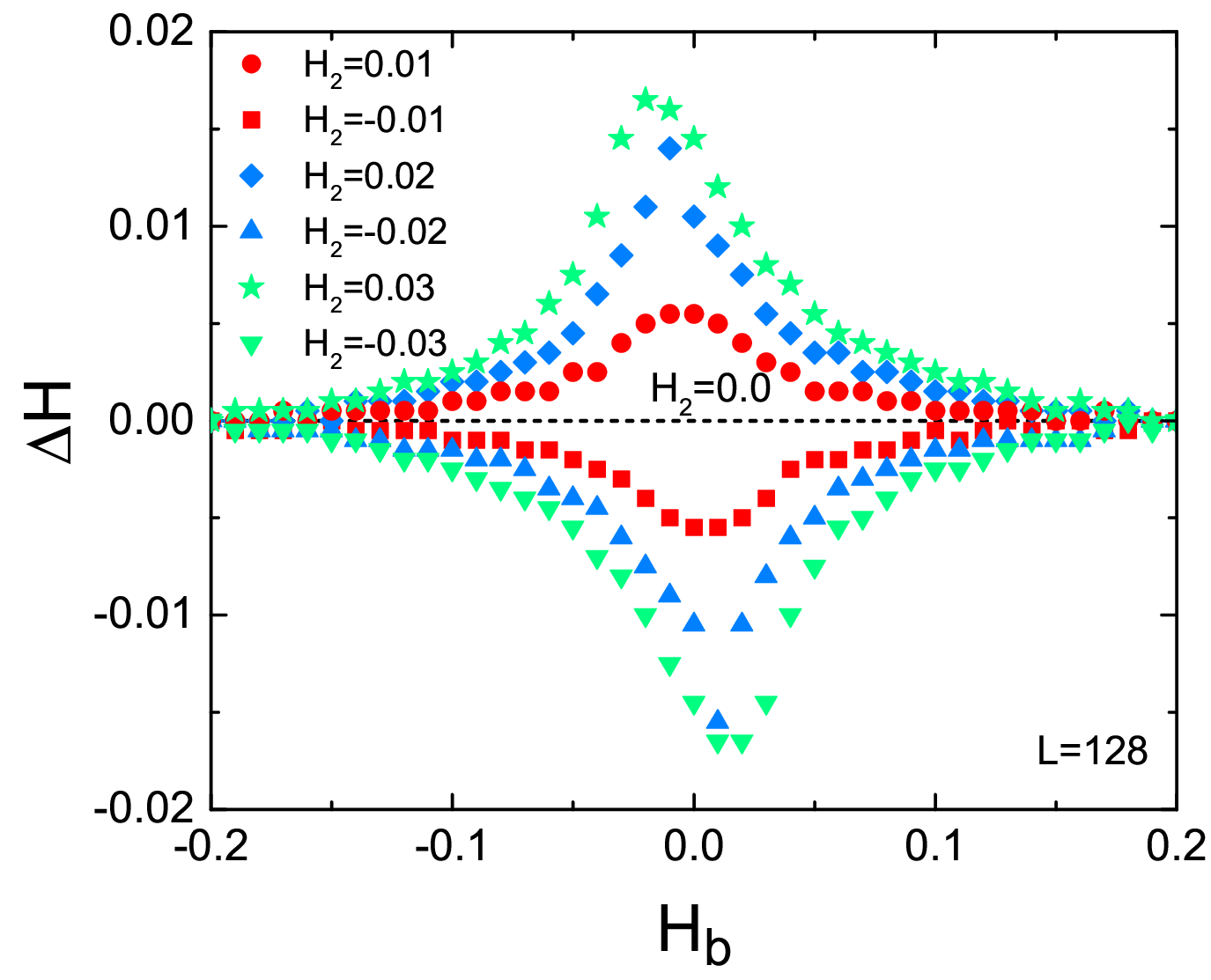}}
	\subfigure[\hspace{0cm}] {\includegraphics[width=7.0cm]{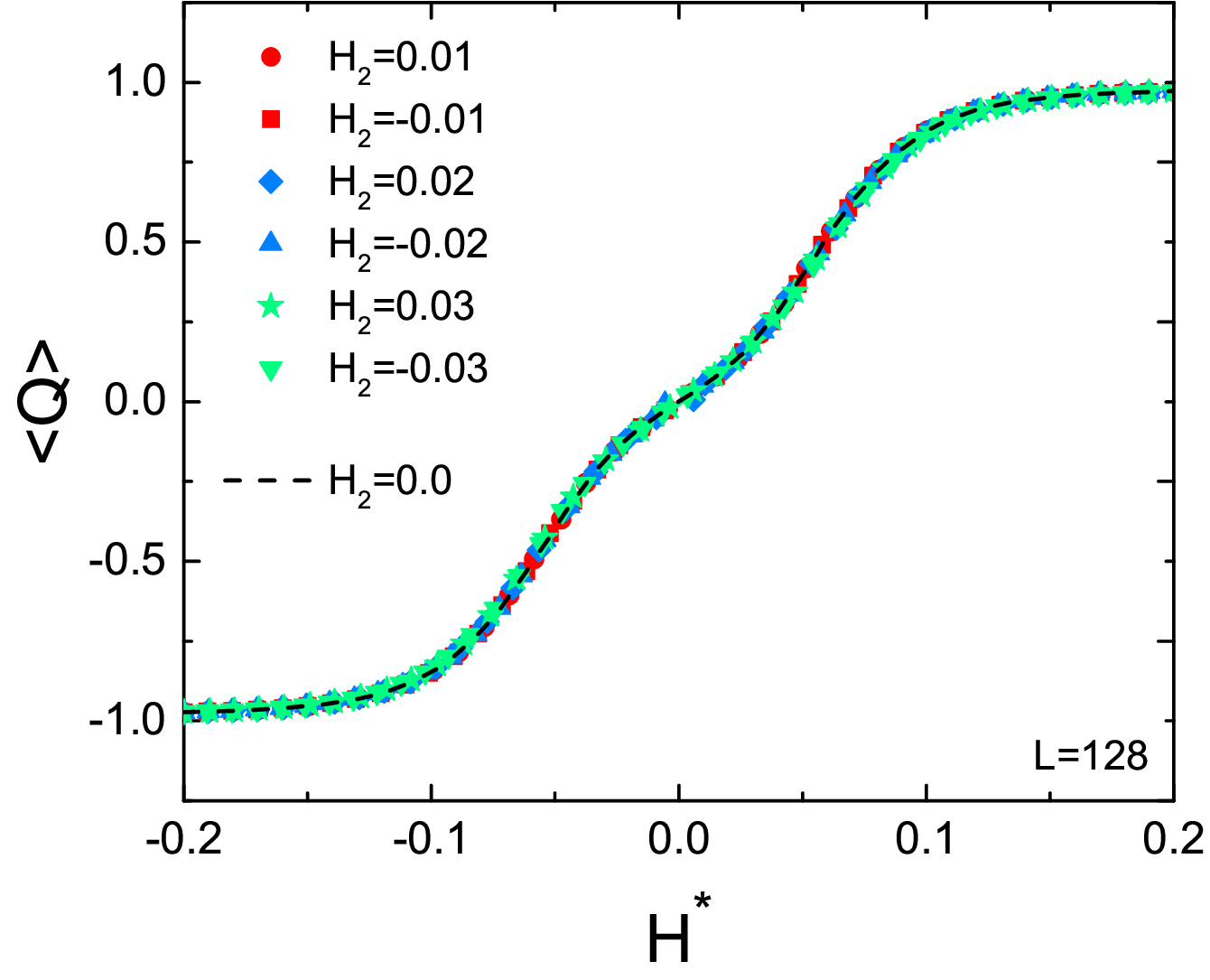}}\\
	\caption{(a) Time dependence of $m(t)$ and $H(t)$ for $H_{0}/J=0.3$, $H_{b}=0.0$ and $H_{2}/J=0.1$. (b) $Q$ versus $H_{b}$ dependence corresponding to dynamic paramagnetic phase with $t_{1/2}=150$ with some selected values of $H_{2}/J=0,\pm0.01, \pm 0.02$ and $\pm0.03$. (c) Variation of $\Delta H$ as a function of $H_{b}$ for the same set of parameters given in (b). (d) $Q$ as a function of $H^{*}$.}\label{fig4}
\end{figure*}

Next, in Fig. \ref{fig4}, we examine the influence of the second harmonic term with amplitude $H_{2}$ and period $P/2$ on the DPT characteristics of the present kinetic Ising system. From Fig. \ref{fig4}a, we see that for nonvanishing values of the amplitude such as $H_{2}/J=0.1$, it is clearly observed   that the half-wave anti-symmetry in $H(t)$ sequence is broken and $\langle Q\rangle$ does not reduce to zero for very large $P$ even when $H_{b}=0.0$. Fig. \ref{fig4}b shows the variation of $\langle Q\rangle$ as a function of $H_{b}$ for some selected values of $H_{2}$ with $P \gg P_{c}$. Note that the property of half-wave anti-symmetry in the dynamic order parameter $Q$, defined as $Q(H_{b})=-Q(-H_{b})$ is only fulfilled when $H_{2}=0.0$. This is due to the fact that this property  is only ensured within a limited regime $(H_{2}=0.0)$ where  $H_{b}$ can be regarded as the conjugate field of $Q$. The amount of the deviation from the half-wave anti-symmetry of $Q$ as a function of $H_{b}$ is depicted in Fig. \ref{fig4}c. It is evident that $\Delta H$ becomes more prominent for larger $H_{2}$ values. Besides, we can observe that the sign of  $\Delta H$ becomes negative when $H_{2}$ is negative, and vice versa. When  $\langle Q\rangle$ is plotted as a function of $H^{*}$ which is calculated with the help of Eqs. (\ref{eq1}) and (\ref{eq2}), we see that the half-wave anti-symmetry property is restored and $\langle Q\rangle$ vs $H^{*}$ curves corresponding to different $|H_{2}|$ values collapse onto each other. It is important to note that the qualitative aspects of these observations are in very well agreement with recently published works \cite{quintana,quintana2}. Hence, we can infer from these results that $H_{b}$ can not be regarded as the conjugate field of $Q$, and the concept of generalized conjugate field $H^{*}$ must be introduced when $H_{2}\neq0$.

\begin{figure}[!h]
	\center
	\includegraphics[width=8.0cm]{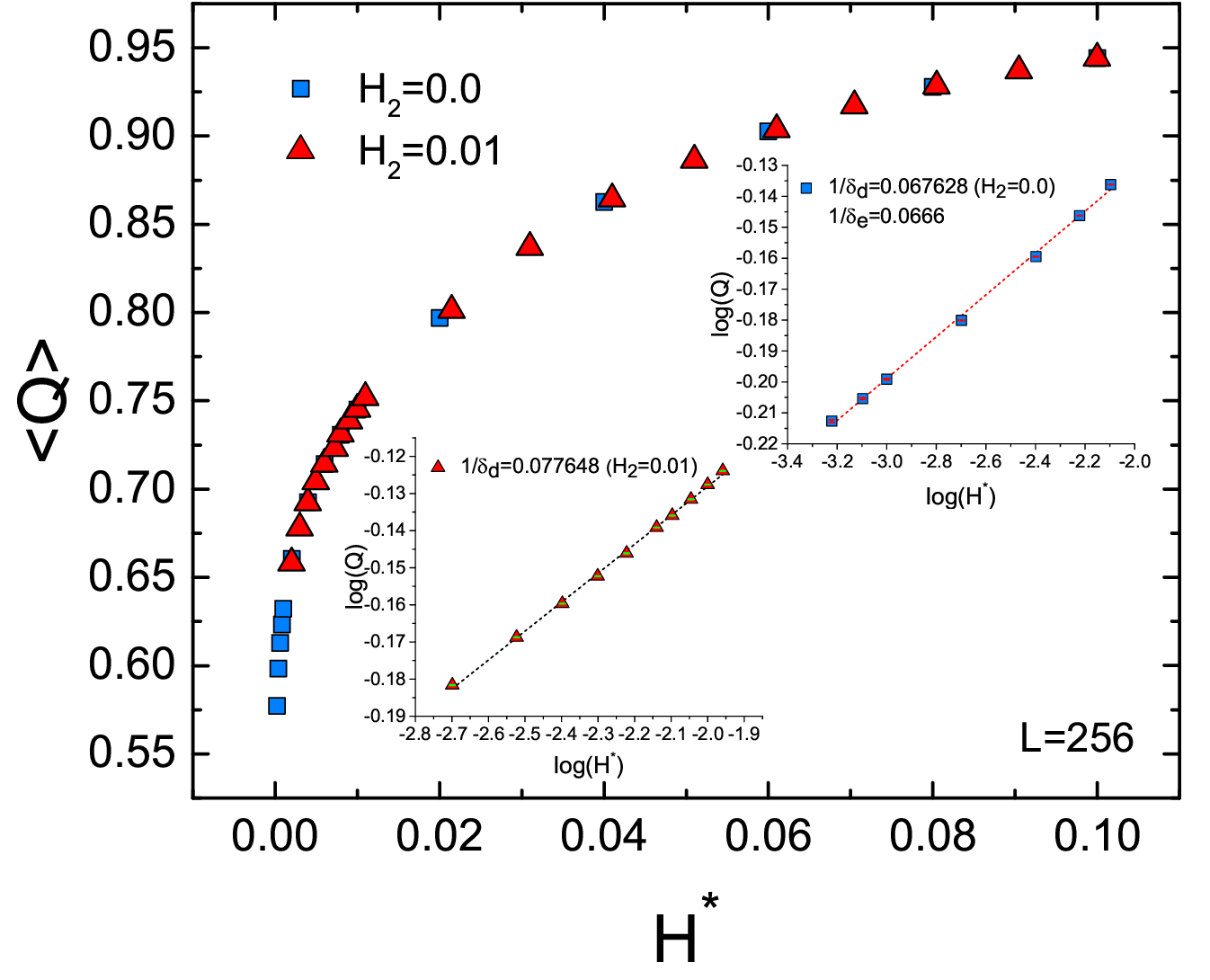}
	\caption{Variation of dynamic order parameter $\langle Q\rangle$ as a function of $H^{*}$. The inset plots show the logarithmic scaling relations calculated according to Eq. (\ref{eq11}) for $H_{2}/J=0.0$ and 0.01.}\label{fig5}
\end{figure}

For the sake of completeness of the present analysis, we can examine whether the  following scaling relation can be verified or not
\begin{equation}\label{eq11}
	Q(H^{*}\rightarrow0)\propto (H^{*})^{1/\delta} \quad \mathrm{for} \ P=P_{c}, \ 
\end{equation}
based on the definition of $H^{*}$ being the conjugate field of $Q$. In terms of the analogy between the DPT and TPT cases, the above scaling relation in Eq. (\ref{eq11}) was previously verified for $H_{2}=0.0$ corresponding to the limiting case $H_{b}=H^{*}$ \cite{yuksel_0} with the associated dynamical scaling exponent $\delta_{d}=14.99$ which is very close to the critical isotherm value $\delta_e=15.0$ of the equilibrium 2D Ising model \cite{mcKenzie}. In Fig. \ref{fig5}, we plot the variation of $\langle Q\rangle$ as a function of non-vanishing $H^{*}$ values for amplitudes of $H_{2}/J=0.0$ and $0.01$ where the curves corresponding to different $H_{2}$ values seem to be overlapping in the entire $H_{2}$ range. Considering a lattice with $L=256$ and for the vanishingly small $H_{2}$ values, we have extracted the critical exponent values according to Eq. (\ref{eq11}) which yields $\delta_d=14.79$ and $\delta_d=12.88$ for $H_{2}=0.0$ and $H_{2}/J=0.01$, respectively. 
Hence, we can conclude that for relatively small, but not vanishing $H_{2}$ values such as $H_{2}=0.01$, the exponent $\delta$ deviates from the corresponding TPT value. 
These results show that although the relations given in Eqs. (\ref{eq1}) and (\ref{eq2}) ensure the proper definition of the generalized conjugate field $H^{*}$, the restoration of similarities between DPT and TPT cases may be failed in terms of the universality aspects related to $H_{2}$. The reason can be explained as follows: Within the framework of mean field theory, it was reported that the presence of small $|H_{2}|$ modifies the critical period $P_{c}$ \cite{quintana}. On the other hand, according to the recent Monte Carlo simulations \cite{vatans}, a dynamic phase transition cannot be observed for sufficiently large $H_{2}$ values, as the Binder cumulant curves for different $L$ do not exhibit a crossing point for $|H_{2}|/J>10^{-3}$. In other words, the dynamic phase transition disappears, and the system attains a dynamically field polarized state given that $|H_{2}|$ is  sufficiently large.     

\section{Conclusion}\label{conclude}
In conclusion, we have examined the recently introduced generalized conjugate field formalism in magnetic systems exhibiting dynamic phase transitions. For this aim, we take 2D kinetic Ising model defined on a honeycomb lattice as a basis, and performed extensive Monte Carlo simulations to identify the critical period at which a dynamic phase transition takes place. Our simulation data shows an
emergent second order dynamic phase transition at $t_{1/2}^{c}=57$ for field sequences exhibiting the half-wave anti-symmetry property for which we have also confirmed that the bias field $H_{b}$ is the conjugate field of the dynamic order parameter $Q$ for conventional sinusoidal magnetic field oscillations.  In the presence of a second magnetic field component with amplitude $H_{2}$ and period $P/2$, the half-wave anti-symmetry is broken and the generalized conjugate field formalism was found to be valid for the present system. However, when we consider log-log plots of $(\langle Q\rangle \ \mathrm{vs} \ H^{*})$ curves for relatively small (in comparison to $H_{0}$), but not vanishing $H_{2}$ values, we found that the scaling exponent significantly deviates from its equilibrium value along with the manifestation of a dynamically field polarized state. Although we have presented simulation results for two-dimensional case, extending the main conclusions of the present work to three-dimensional lattices is straightforward. In this regard, we hope our work will shed some light on the emergent 
follow-up works on the theory of dynamic phase transitions.

%\section*{Appendix}

\begin{acknowledgments}
The computational resources are provided by TUBITAK ULAKBIM, High Performance 
and Grid Computing Center (TR-Grid e-Infrastructure). The author would like to thank A. Berger from CIC nanoGUNE for fruitful discussions on the generalized conjugate field analysis in kinetic Ising models.
\end{acknowledgments}

\newpage
\bibliography{references}

\end{document}